\documentclass[pdflatex,sn-mathphys-num]{sn-jnl}

\usepackage[mathlines]{lineno} 
\usepackage{graphicx}%
\usepackage{multirow}%
\usepackage{mathtools}
\usepackage{amsmath,amssymb,amsfonts}%
\usepackage{amsthm}%
\usepackage{mathrsfs}%
\usepackage[title]{appendix}%
\usepackage{xcolor}%
\usepackage{textcomp}%
\usepackage{manyfoot}%
\usepackage{booktabs}%
\usepackage{algorithm}%
\usepackage{algorithmicx}%
\usepackage{algpseudocode}%
\usepackage{listings}%
\usepackage[T1]{fontenc}
\usepackage{xcolor}
\usepackage{pdfpages}
\usepackage{soul}
\soulregister\ref7
\soulregister\cite7
\usepackage[normalem]{ulem}

\theoremstyle{thmstyleone}%
\theoremstyle{thmstyletwo}%

\theoremstyle{thmstylethree}%

\begin{document}

\title[Article Title]{Polarization-independent deterministic mode localization in a photonic lantern}

\author[1]{Harikumar K Chandrasekharan*}
\author[1]{Ross Donaldson}

\affil[1]{Scottish Universities Physics Alliance, Institute of Photonics and Quantum Sciences, School of Engineering
and Physical Sciences, Heriot-Watt University, David Brewster Building, Edinburgh EH14 4AS, Scotland,
UK}

\affil[*]{hk47@hw.ac.uk}


\abstract{Coherent interference in multimode photonic systems underpins scalable, high-fidelity control for beam shaping, power delivery, and signal processing, yet most existing approaches rely on bulky adaptive optics or polarization-sensitive waveguides. Here, we demonstrate an all-fiber, polarization-independent coherent mode-recombination scheme that deterministically localizes Gaussian-like spots—with a Gaussian similarity index up to 0.95—at three distinct positions on the multimode facet of a commercial three-mode graded-index photonic lantern (PL). The device coherently combines the lantern’s individual outputs using piezoelectric phase shifters and a reciprocal Faraday-mirror feedback loop, which enforces polarization reciprocity and eliminates alignment sensitivity. This configuration achieves near-unity (100\%) relative mode-conversion efficiency, three-spot switching, and long-term stability with sub-micron centroid drift (0.55~$\mu\mathrm{m}$) without active feedback. The phase-locked profiles maintain high Gaussian correspondence, strong spatial confinement, and high single-mode coupling efficiency, demonstrating robustness under laboratory-scale perturbations. Numerical simulations quantitatively reproduce the experimental recombination dynamics and further establish scalability through six-mode commercial-lantern modeling. The polarization-insensitive, compact, and low-loss architecture establishes PLs as practical engines for coherent beam forming and deterministic spatial localization, enabling turbulence-resilient beam delivery, reconfigurable mode-division multiplexing, biomedical imaging and sensing, and quantum photonics, while reducing system complexity and preserving efficiency.}




\maketitle

\section{Introduction}\label{sec1}

Optical fibers are fundamental components in modern photonic systems, enabling the transmission of optical signals across a wide range of applications, including long-haul optical communications\cite{doi:10.1049,Fatome:10}, quantum key distribution\cite{articleqkd1,20.044006}, and imaging\cite{science.abl3771,Li:00}. While single-mode fibers (SMFs) enable higher bandwidth for signal transmission\cite{article222}, their small core size requires precise alignment, introducing instrumental complexity\cite{1128517}. One approach to improve coupling efficiency is to use larger-core multimode fibers (MMFs)\cite{601675,9277524}. However, a fundamental limitation of MMFs in signal transport applications is modal dispersion, arising from the complex propagation dynamics of light within the larger core\cite{Kovaevi2017ARA,2010xzc}. Despite the challenges of signal coupling, modern optical communication systems continue to favor SMFs for their high bandwidth capacity and minimal signal dispersion. An ideal optical fiber system would combine the high multiplexed coupling efficiency of MMFs with the bandwidth and signal-processing advantages of SMFs. Significant research efforts have focused on mitigating modal dispersion and signal broadening in MMFs through advanced wavefront shaping\cite{articlebnm,articlevgf,1.5136334,9420070} and adaptive optics-based beam combination techniques\cite{MORI2013132,articlecdt,Castillo:25}. However, despite these advances, achieving MM light coupling and SM conversion in waveguide systems still relies on complex adaptive optical components\cite{Chen:15,refId0,Chen:24,Zhou2025_HPL}.

Conventional coherent beam-shaping and mode-recombination techniques mentioned above often rely on adaptive optics or spatial light modulators, which require active wavefront sensing, iterative feedback, and in some devices, precise polarization alignment\cite{articlebnm,articlevgf,1.5136334,9420070,MORI2013132,articlecdt,Castillo:25,Chen:15,refId0}. While powerful, these systems are bulky, alignment-sensitive, and prone to instability when environmental conditions fluctuate. Their complexity and limited scalability make them impractical for compact or fiber-integrated architectures, where passive reciprocity and long-term stability are critical. Overcoming these challenges calls for a self-referenced, polarization-independent approach capable of deterministic mode control without external wavefront correction.

With the growing demand for scalable coherent beam combining and multimode-to-single-mode (MM–SM) conversion, hybrid architectures offer compelling advantages for applications requiring efficient light collection and precise modal control. 
Such systems are particularly relevant in coherence-critical domains such as free-space communication and quantum key distribution (QKD), where maintaining SM performance is essential to minimize modal dispersion and preserve timing precision in photon detection. 
A promising approach for achieving efficient MM light collection and reformatting into SM outputs is the use of photonic lanterns (PLs)\cite{Leon-Saval:10,Birks:15}, which provide low-loss, unitary mapping between MM and multiple SM channels.
With a multimode input and multiple single-moded (or few-moded) outputs, PLs have proven to be excellent candidates for multiplexed light collection, detection, and processing, having already been demonstrated for low-order mode wavefront sensing\cite{articleNorris} and optical beam-steering \cite{Cruz-Delgado:21}. 

In addition to their established roles in time-stretch imaging\cite{articleHKC}, adaptive spatial mode control\cite{articlePLMC}, compressive imaging\cite{articleCD}, space- and mode-division multiplexing\cite{Chen:23,Sai:17,articleAmado}, astronomical instrumentation\cite{2012MNRAS,Lin:25}, and optical coherence tomography\cite{articleOCT}, PLs are now emerging as key enablers for high-performance QKD receivers and as promising low-complexity alternatives to adaptive optics in free-space optical links\cite{Billault:25,beraza2025,Chandrasekharan:24}. While PLs have demonstrated impressive capabilities for MM light collection and coherent beam reformatting, their potential for deterministic, reciprocal, and dynamically controllable beam shaping remains largely unexplored. Unlocking this capability would elevate PLs from passive mode converters to active, stable, and compact engines for coherent MM control, offering a powerful substitute for bulky and power-hungry adaptive optical systems.

\begin{figure*}[b]
		\centering
		{\includegraphics[width=\linewidth]{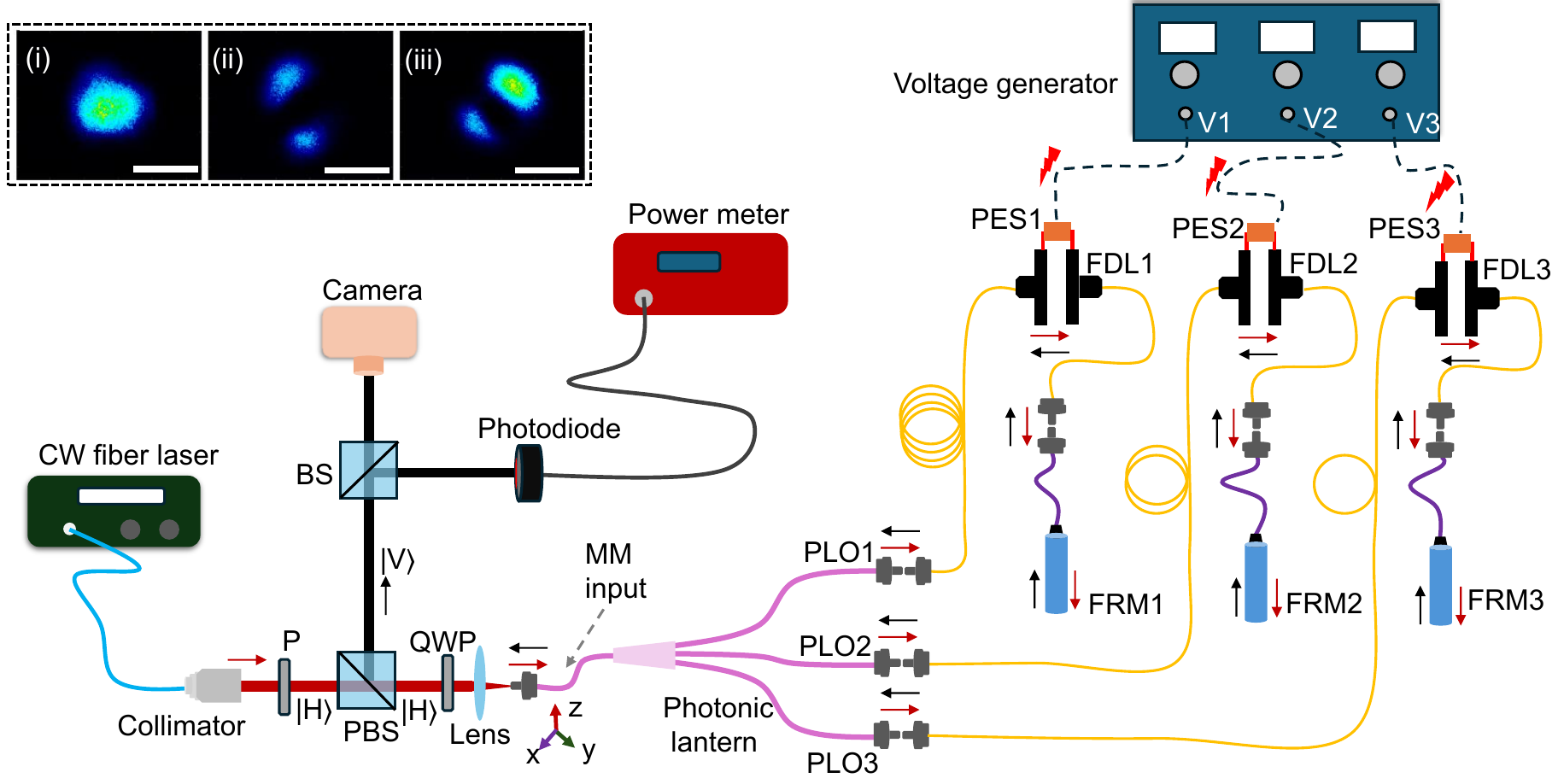}}
		\caption{Experimental system used for the coherent mode recombination system. The red arrow denotes the forward-propagating optical signal, whereas the black arrow indicates the trajectory of the back-reflected light. Inset - PL mode profiles obtained at the MM input when selectively coupling light from the PL output fibers, (i) LP$_{01}$ mode, (ii) LP$_{11a}$ mode, (iii) LP$_{11b}$ mode. Scale bar, 0.5 mm. P – polarizer, PBS – polarizing beamsplitter, BS- beamsplitter, QWP – quarter waveplate,  PLO- photonic lantern output, FDL- fiber delay line, PES- piezoelectric stack, FRM- Faraday rotating mirror.}
	\label{fig1}
	\end{figure*}
    
In this work, we introduce, to our knowledge, the first all-fiber, polarization-independent coherent mode-recombination scheme that deterministically localizes Gaussian-like spots at multiple positions across the MM facet of a PL using only phase control and Faraday-mirror reciprocity, and requiring no complex adaptive optics or spatial light modulators. Our approach employs piezoelectric (PE) phase shifters together with a reciprocal Faraday-reflecting loop to impose precise and stable control over the relative modal phases, achieving near-unity (100\%) image-based mode-conversion efficiency in a commercial three-mode PL. By tuning only the PE-driven phase differences, we reproducibly steer and lock Gaussian-like interference profiles to arbitrary locations within the MM core. We further reproduce the three-mode experimental dynamics numerically and demonstrate scalability via six-mode commercial PL simulations. This establishes a new operating regime for PL as coherent, deterministic MM controllers rather than passive mode couplers—offering a scalable, alignment-free, and inherently polarization-independent method for MM control without complex adaptive optics. The demonstrated mechanism provides a robust platform for coherent beam shaping, optical signal processing, and next-generation photonic communication technologies.

\section{Results}\label{sec2}
\textbf{Experimental set-up:} A schematic of the experimental setup used to demonstrate the deterministic mode localization is shown in Figure~\ref{fig1}. A fiber-coupled coherent continuous-wave (CW) laser (DFB1550P, Thorlabs) operating at 1553 nm (0.05 nm FWHM bandwidth, coherence length $\approx$48 mm) was collimated and directed to a polarizing beamsplitter (PBS) after passing through a polarizer (P). The transmitted beam was then coupled into the MM input of the PL using a focusing lens. A quarter-wave plate (QWP) inserted before the coupling lens allowed controlled adjustment of the input polarization state, enabling verification of the polarization insensitivity of the mode-recombination process.

 The PL used for the coherent mode combination demonstration, inherently polarization insensitive due to its symmetric and guided-wave design, was commercially obtained from Phoenix Photonics Ltd (Part number: 3PLS-GI-15-1-1-1-MS). The PL was designed to support three distinct spatial modes: the fundamental LP$_{01}$ mode and the higher-order LP$_{11a}$ and LP$_{11b}$ modes (see inset images in Figure \ref{fig1}). Each output mode was supported over a length of 1 m, fabricated from SMF-28 fiber. These outputs are indicated by PLO1, PLO2, and PLO3 corresponds to LP$_{01}$, LP$_{11a}$ and LP$_{11b}$ respectively in Figure \ref{fig1}. The MMF at the output was fabricated from a two-mode graded-index fiber with 20~$\mu$m core diameter. The PL is characterized in terms of insertion loss and numerical aperture (NA) and is presented in Supplementary Notes 1\&2.  

 \begin{figure*}[t]
		\centering
		{\includegraphics[width=\linewidth]{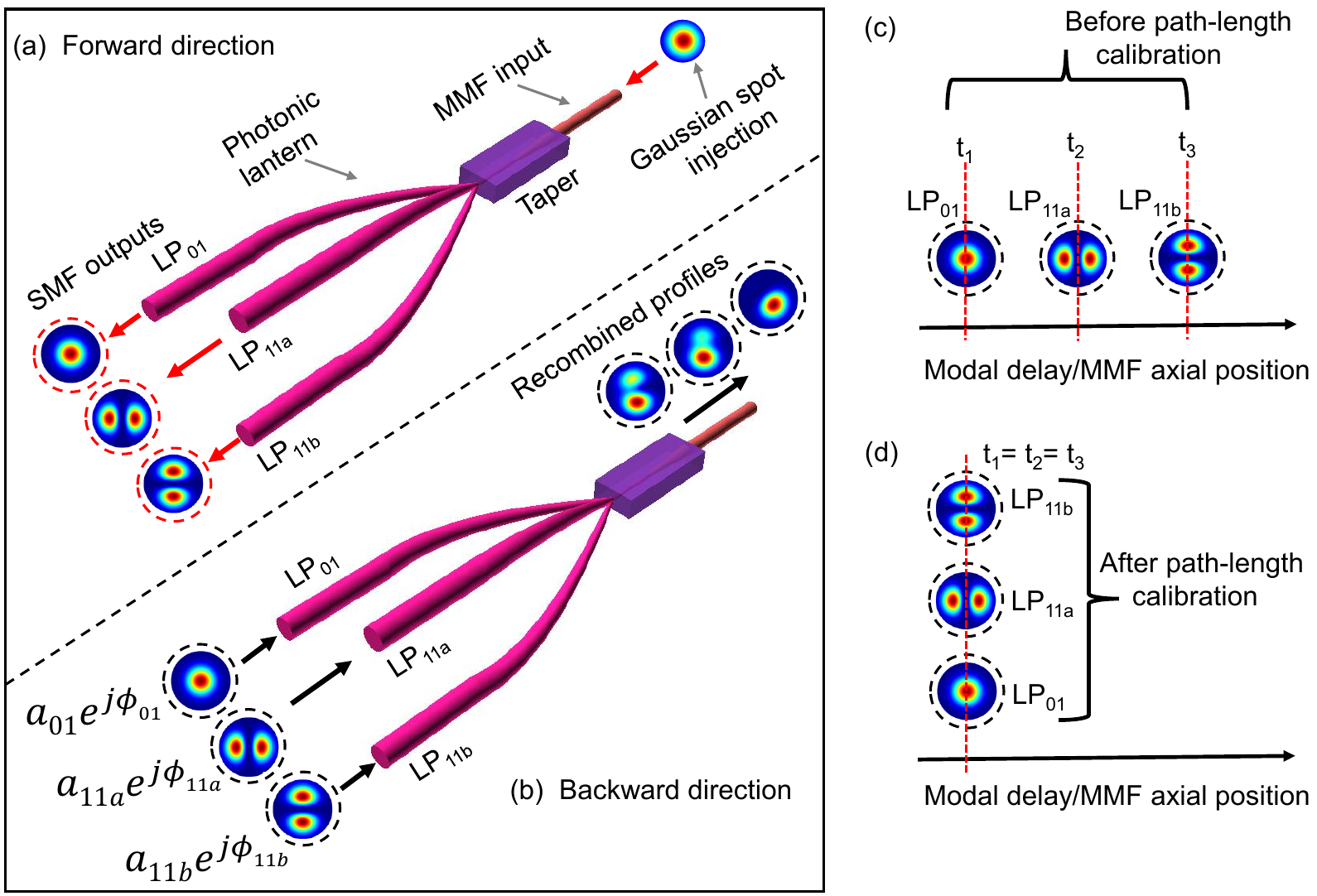}}
		\caption{Forward and backward light propagation through the PL and path-length calibration. (a) A Gaussian input excites the supported LP modes in controlled proportions. (b) After reflection from the Faraday reflectors, the three returned modal fields recombine coherently in the MMF core, producing a distinct interference pattern at the MMF output facet based on the relative modal delays/phases (see Supplementary Material). The terms $a_{\ell m}$ and $\phi_{\ell m}$ denote the amplitude (the power weighting of each LP mode after propagation) and phase (accumulated along the path and upon reflection from the FRM) of mode $LP_{\ell m}$, respectively. (c) When the optical path-lengths of the PL arms are not equalized, the modal delays of the returned modes are mismatched, resulting in modes arriving at different axial locations along the MMF with different arrival times (t$_{1}$-t$_{3}$). (d) After path-length calibration of the PL arms, the reflected modes arrive at the same time and axial location in MMF.}
	\label{fig2}
	\end{figure*}

To demonstrate the state-of-the-art, fiber delay lines (FDLs) and fiber Faraday reflecting mirrors (FRMs) were incorporated at each PL output with equal lengths, enabling both reflection from the FRMs and precise phase control of the back-reflected light. The FDLs were configured to allow adjustment of path length by varying the relative positions of the input and output fiber facets using a linear translation stage with a travel range of 60 mm. In our implementation, fine phase modulation was achieved by introducing PE actuators within the delay lines, where small path length variations were controlled by applying voltage signals to the PE stacks. Three PE stacks were used to independently adjust the optical path length—and thus the phase—at each of the three PL outputs. 

The first step toward achieving efficient mode recombination is a coarse calibration of the path lengths in the PL. After light is injected into the MMF input, the optical power is distributed among the three output ports, as conceptually represented in Figure~\ref{fig2}a, and the reflected signals return through the respective arms, after the FDLs and FRMs, carrying their corresponding modal fields\cite{articlePLMC}, as shown in Figure~\ref{fig2}b. In an ideal configuration where all fiber segments and components in the PL arms have identical optical lengths, the returned modes would arrive at the MMF core simultaneously and recombine to form the desired mode profiles under phase control. In practice, however, each PL output port and its associated components introduce a distinct modal delay. Consequently, the returned modes reach the MMF facet at different times and spatial locations, as illustrated in Figure~\ref{fig2}c, which prevents coherent recombination. When the path-length calibration is performed, these modal delays are equalized, enabling the returned modes to arrive synchronously at the MMF input without relative modal delays, as represented in Figure~\ref{fig2}d. 

Note that Figure~\ref{fig2}b also includes simulated representations of three modal recombination scenarios—two phase-mismatched (delay-offset) cases and one phase-matched case. These simulations are intended as idealized illustrations of how modal interference manifests when the three guided modes of the lantern are coherently recombined. The model assumes a perfectly mode-selective three-mode PL supporting the canonical LP$_{01}$, LP$_{11a}$, and LP$_{11b}$ spatial mode profiles of a weakly guiding MMF, and does not incorporate wavefront distortion, component-induced aberrations, or fiber inhomogeneity. It therefore illustrates an idealized modal superposition rather than a full physical model of the experimental system (see Supplementary Material for complete simulations).

To measure the modal delays and to perform the path-length calibration, a pulsed 1550-nm laser (PicoQuant Laser) with a limited coherence length was coupled into the MM end of the PL and monitored the light that was reflected. The reflected signals were detected using an InGaAs single-photon avalanche diode (SPAD) (id220-FR, ID Quantique) placed in the detection arm. By blocking all but one PL output at the free-space delay line, the return from each PLO was isolated, and its temporal histogram was recorded using a time-tagger (ID900, ID Quantique) operating in time-correlated single-photon counting (TCSPC) mode with a 100-ps bin resolution. This process provided clear timing information for each PLO signal peak, allowing us to manually adjust the free-space delay line so that all returned signals fell within the coherence length of the CW laser used later for recombination.

For the PL used in this work, the TCSPC arrival traces for the LP$_{01}$, LP$_{11a}$, and LP$_{11b}$ modes appeared within a 300-ps window, which is approximately twice the coherence time corresponding to the CW laser’s coherence length in silica fiber (coherence length $\approx$33 mm, equivalent to $\approx$160~ps propagation time). To ensure that all modal peaks lay within the laser coherence interval, only the LP$_{01}$ trace was shifted by introducing an additional delay ($\approx$35~mm) at the free-space fiber delay line corresponding to two TCSPC bins ($\approx$200~ps), while no adjustment was applied to the LP$_{11a}$ and LP$_{11b}$ paths. Further details of the calibration procedure and TCSPC acquisition are provided in Supplementary Note~3.

Following coarse path-length tuning, the pulsed source was replaced with the high-coherence CW laser (DFB1550P) to ensure stable interference during phase control. PE stacks were then incorporated into the delay lines as shown in Figure~\ref{fig1} to enable fine phase modulation of the optical signals, facilitating precise mode recombination at the MM core. The PE stacks, designed to achieve a maximum displacement of 150 $\mu$m, were used to modulate the phase of the incoming light as a function of the applied voltage. A series of voltage values was applied to the PE stacks in a programmable sequence, inducing controlled phase shifts across the PL output delay lines. These phase shifts produced varying interference conditions that were highly sensitive to even small changes in applied voltage. The phase-induced interference effects were monitored on-the-fly using a 1550 nm camera (WinCamD-LCM, DataRay) positioned at the reflection arm of the PBS. To fully capture the mode profile changes at the MM end, the voltage sequence was set to update every 1 second, while the camera was operated at a frame rate of 5–10 frames per second. The entire process—including on-the-fly voltage adjustments, corresponding power meter readings on a power meter, and image capture—was recorded using the FastStone Capture screen recorder software. This ensured that the voltage settings and corresponding interference patterns could be replicated, which is essential for future mode combination strategies.\\
\begin{figure*}[t]
		\centering
		{\includegraphics[width=\linewidth]{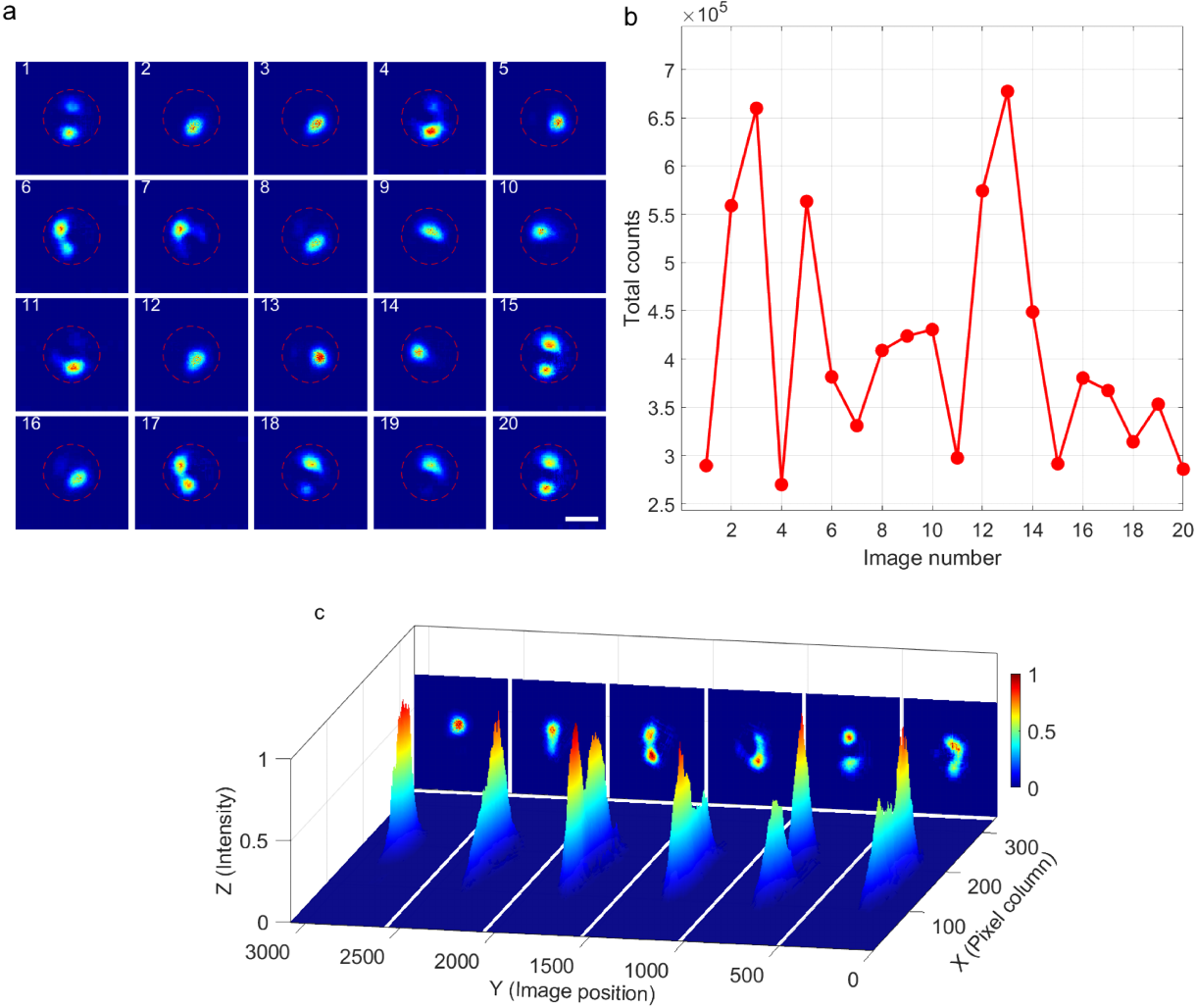}}
		\caption{(a) 20 random mode interference profiles obtained at the MM end of the PL for various piezo voltage combinations. The frames are extracted from the on-the-fly video file during the voltage scan. As seen, the mode energy is concentrated in Gaussian-like profiles on various voltage combinations. The dotted circle outlines the approximate core region of the MM end. All images cover the same spatial area, scale bar, 0.5 mm. (b) Respective mode energies in the image frames, quantified by summing the pixels within the core region. See Supplementary material (Figure S7) for simulated recombined profiles and efficiencies. (c) 3D visualization of 6 random images with smoothed surface rendering and corresponding images projected on the YZ wall, capturing the diverse spatial evolution of the light field upon PE scan. The final frame exhibits a Gaussian-like intensity distribution, indicative of effective MM beam shaping and high-fidelity conversion to a desired SM spatial profile.}
	\label{fig3}
	\end{figure*}
    
\noindent\textbf{Deterministic three-spot mode recombination:} After path length calibration and finalization of voltage scanning parameters, the scan is performed by sequentially sweeping the voltages across the PE stacks at the PL delay lines. Light was injected into the MM end of the PL, and the relative output intensities were measured as 46\%, 29\%, and 25\% for PLO1, PLO2, and PLO3, respectively. This relative energy distribution was quantified by selectively measuring the signal intensities on a power meter positioned at each of the PL outputs. A MATLAB script interfaced with the voltage generator to drive the PE stacks, producing controlled phase shifts in the back-reflected light from the PL outputs terminated with FRMs.

Images captured by the camera were recorded in real-time with FastStone screen recorder, and individual frames were subsequently extracted to quantify the phase-scanned mode recombination. Figure \ref{fig3}a displays 20 representative random image frames (corrected for background noise) extracted from a continuous voltage scan, selected to show the diversity of optical field distributions resulting from dynamic phase modulation at the PL output. These frames reveal a broad range of interference conditions between the spatial channels, including multiple instances of constructive and destructive interference. Notably, certain frames exhibit strong spatial localization with Gaussian-like intensity profiles, indicating coherent mode superposition with high modal overlap. Such energy localization enables precise and efficient MM-to-SM conversion, which is particularly beneficial for single-photon QKD receivers, where spot size directly impacts the achievable key rate\cite{qtc2.12091}.

To provide a quantitative assessment of the interference dynamics, the total photon count in each frame was computed by summing the pixel intensities within the defined MM core region. The resulting values, plotted in Figure \ref{fig3}b, exhibit significant fluctuations corresponding to the evolving interference states. Peaks in the total count trace denote constructive interference events with efficient modal combination, while dips indicate destructive interference and phase mismatch between channels. This analysis underscores the system’s capacity for real-time phase-sensitive control over mode interference patterns at the output facet.

Figure \ref{fig3}c shows smoothed 3D surface plots for six representative output intensity profiles from the PL, acquired during a voltage sweep of the PE stacks. Each plot is normalized and accompanied by the corresponding 2D intensity projection on the yz-plane, illustrating the evolving interference patterns at the MM output as the phase of the back-reflected signal is tuned via the fiber-integrated Faraday reflector. The surface profiles highlight the spatial evolution of mode combinations across the lantern output, revealing how varying voltage values modulate the resultant intensity structure. This figure visually demonstrates the platform's ability to achieve controlled and reconfigurable modal outputs through all-fiber, polarization-independent tuning.

 \begin{figure*}[t]		\centering{\includegraphics[width=\linewidth]{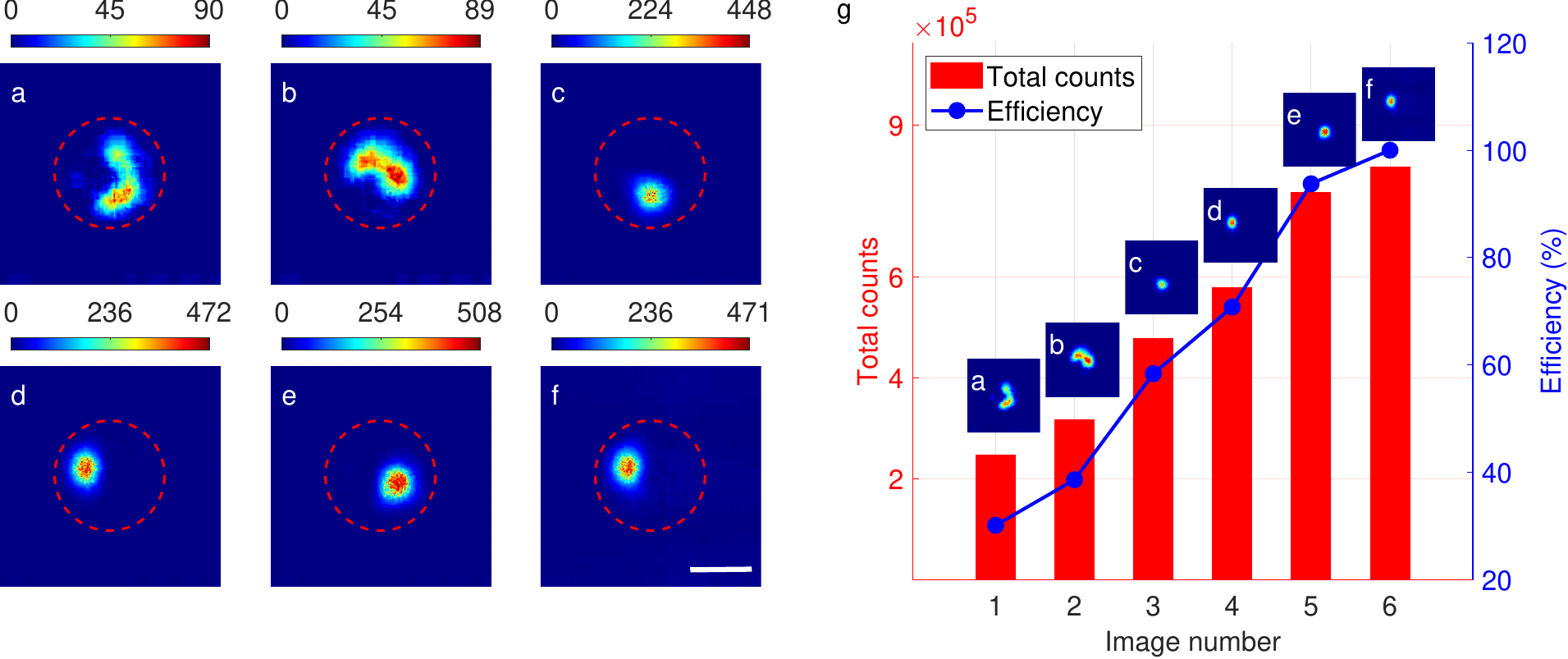}}
		\caption{ Representative mode-interference outcomes and efficiency metrics. (a-f) Mode intensity profiles extracted from the recorded video sequence, illustrating key interference outcomes. Subfigure (a) presents the image frame with the minimum photon flux, subfigure (b) highlights a highly non-Gaussian interference state, and subfigures (c-f) present image frames with distinct instances of Gaussian-like mode interference patterns (see Supplementary material for simulated 3-spot generation). Subfigure (f) is the mode profile associated with the maximum photon flux. Scale bar, 0.5 mm. (g) Total photon counts and efficiencies of the generated mode profiles. Bar plots represent total photon counts for each of the six frames (subfigures (a–f), overlaid on the bars), calculated by integrating pixel counts within the defined MM core region. The line plot (blue) shows the energy conversion efficiencies of the mode profiles, calculated by normalizing their photon fluxes to that of subfigure (f), which exhibits the highest photon flux in the entire video.}
	\label{fig4}
	\end{figure*}

A key novelty of our experimental system is its ability to dynamically position Gaussian-like mode profiles at different locations across the MM end, as presented in Figure\ref{fig4}. In particular, three distinct Gaussian-like localization states were obtained (Fig. \ref{fig4}c–e), with image-based conversion efficiencies of 60\%, 95\%, and 100\%. Here, the “conversion efficiency” is defined purely from the recorded camera frames: for each frame, we compute the total photon counts integrated over the entire MM-core image. The frame exhibiting the maximum total counts across the full measurement sequence is taken as the 100\% reference, and the efficiencies of all other localization states are expressed as a percentage of this maximum. In our measurements, the highest-efficiency state coincides with the frame of maximum photon flux, yielding a relative efficiency of 100\%.

This image-based definition provides a self-consistent and interference-synchronous measure of the recombined optical power, as the camera frames capture the full spatial distribution of the field at the exact phase state being generated. Although our setup and screen-capture software simultaneously record the absolute power-meter readings, we rely on the camera-based metric because it directly reflects the modal recombination quality, avoids the latency associated with meter sampling, and enables reliable comparison across rapidly varying interference states. Absolute efficiencies can still be obtained by locking any chosen mode profile, maintaining its phase state, and reading out the corresponding power-meter value through a straightforward extension of the control code. In our case, to confirm that the camera counts provide a quantitative proxy for optical power, we independently calibrated the integrated camera response against a reference power meter (Supplementary Note 4). This calibration establishes the camera as a calibrated photodetector, allowing any recorded interference frame to be converted into an absolute power value and its associated recombination efficiency.

The variation in efficiency among localization states arises from the controlled modal interference within the PL (see Methods and Supplementary material). In the highest-efficiency state (100\%), phase-controlled recombination coherently concentrates the higher-order modal power (54\%) together with the fundamental-mode component (46\%) into a single localized spot. For other localized states, although similar Gaussian confinement was achieved, reduced efficiencies result from specific phase configurations that induce destructive interference among the contributing modes, thereby diminishing the total output intensity.\\
\begin{figure*}[t]
		\centering{\includegraphics[width=\linewidth]{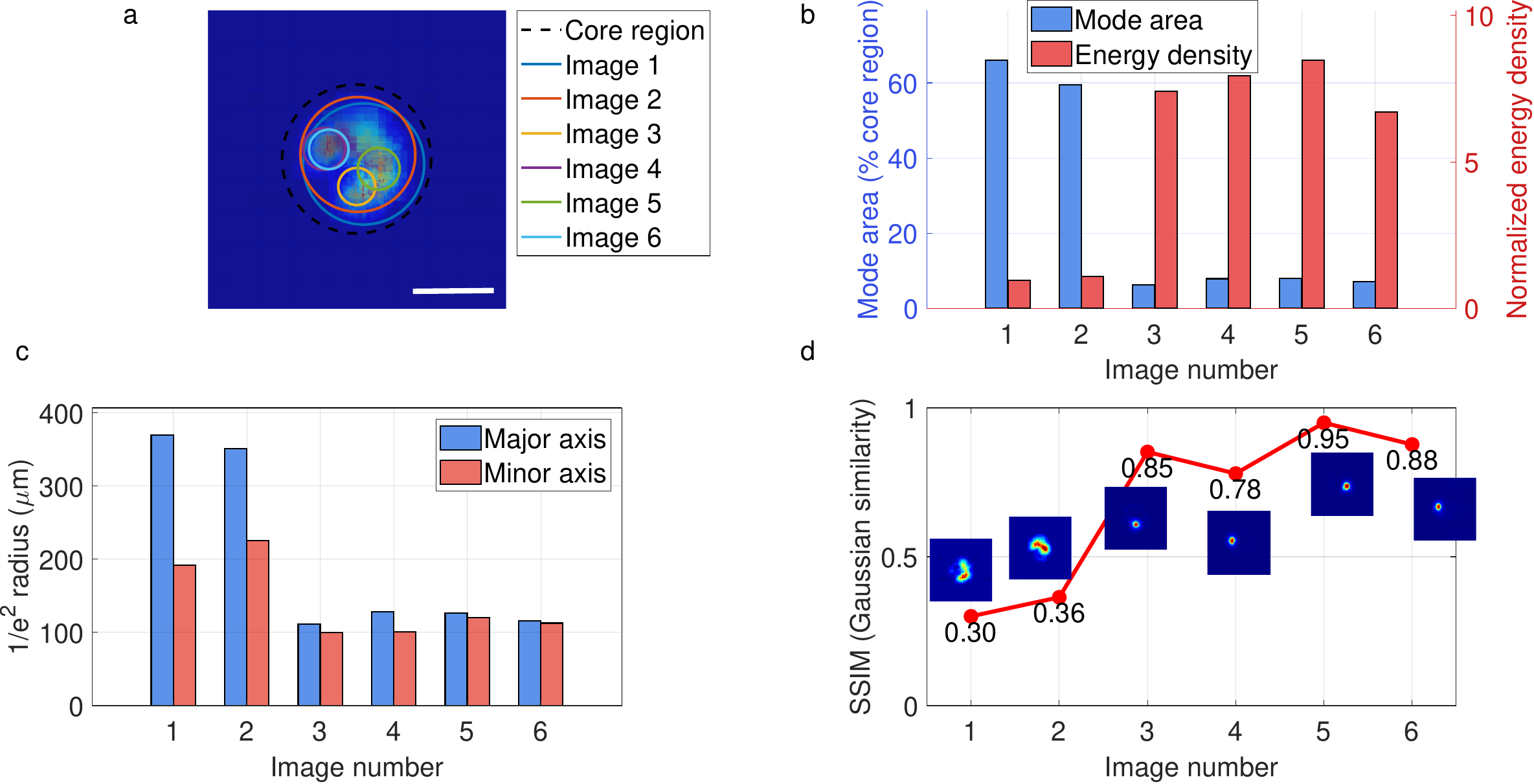}}
		\caption{Spatial mode confinement, energy distribution, and Gaussian similarity of the mode profiles presented in Figure~\ref{fig4}. (a) Measured intensity overlays with 1/e$^{2}$ contours showing occupied modal areas of the images within the MM fiber core. Scale bar, 0.5 mm. (b) Comparison of mode area (blue) and normalized energy density (red), highlighting confinement. (c) Major and minor 1/e$^{2}$ radii representing Gaussian similarity for each mode. (d) SSIM between measured and ideal Gaussian profiles with representative beam thumbnails.}
	\label{fig5}
	\end{figure*}
    
\noindent\textbf{Spatial mode confinement and Gaussian similarity analysis:} In coherent mode recombination, spatial confinement is critical to maintain phase coherence and ensure efficient constructive interference among modes. Conventional approaches, such as those employing adaptive optics or free-space beam combining, often suffer from modal cross-coupling and imperfect spatial overlap, limiting beam quality. We quantitatively analyzed the spatial confinement and Gaussian similarity of the generated modal profiles to characterize modal purity and interference efficiency across the recombined modes. As shown in Figure~\ref{fig5}, the spatial mode confinement and Gaussian similarity analysis reveal beam quality across the six recorded mode profiles. Images~1–6 in Figure~\ref{fig5} correspond to panels~(a–f) in Figure~\ref{fig4}, allowing direct comparison between the measured intensity patterns and their quantitative confinement characteristics. 

To characterize the spatial profiles, the 1/e$^{2}$ beam widths were obtained along the principal axes of each measured mode, and corresponding symmetric Gaussian references were generated for comparison. The overlaid 1/e$^{2}$ contours in Figure~\ref{fig5}a indicate that the modes in Images~3–6 are confined within the MM core, exhibiting smaller spatial spread compared to images~1 and~2. This trend is quantitatively supported in Figure~\ref{fig5}b, where the occupied core area decreases while the normalized energy density rises, confirming stronger modal confinement. The second-moment analysis in Figure \ref{fig5}c shows that images 1 and 2 exhibit a pronounced mismatch between the major and minor 1/e$^{2}$ radii, while images 3–6 display closely matched radii, indicative of more circular and symmetric beam shapes. Consistently, the Gaussian structural similarity index (SSIM) for the localized spots in Figure~\ref{fig5}d ranges from $0.78$ to $0.95$ (images 3-6), corroborating their improved Gaussian correspondence and enhanced spatial localization within the core.\\
\begin{figure*}[t]		
\centering{\includegraphics[width=\linewidth]{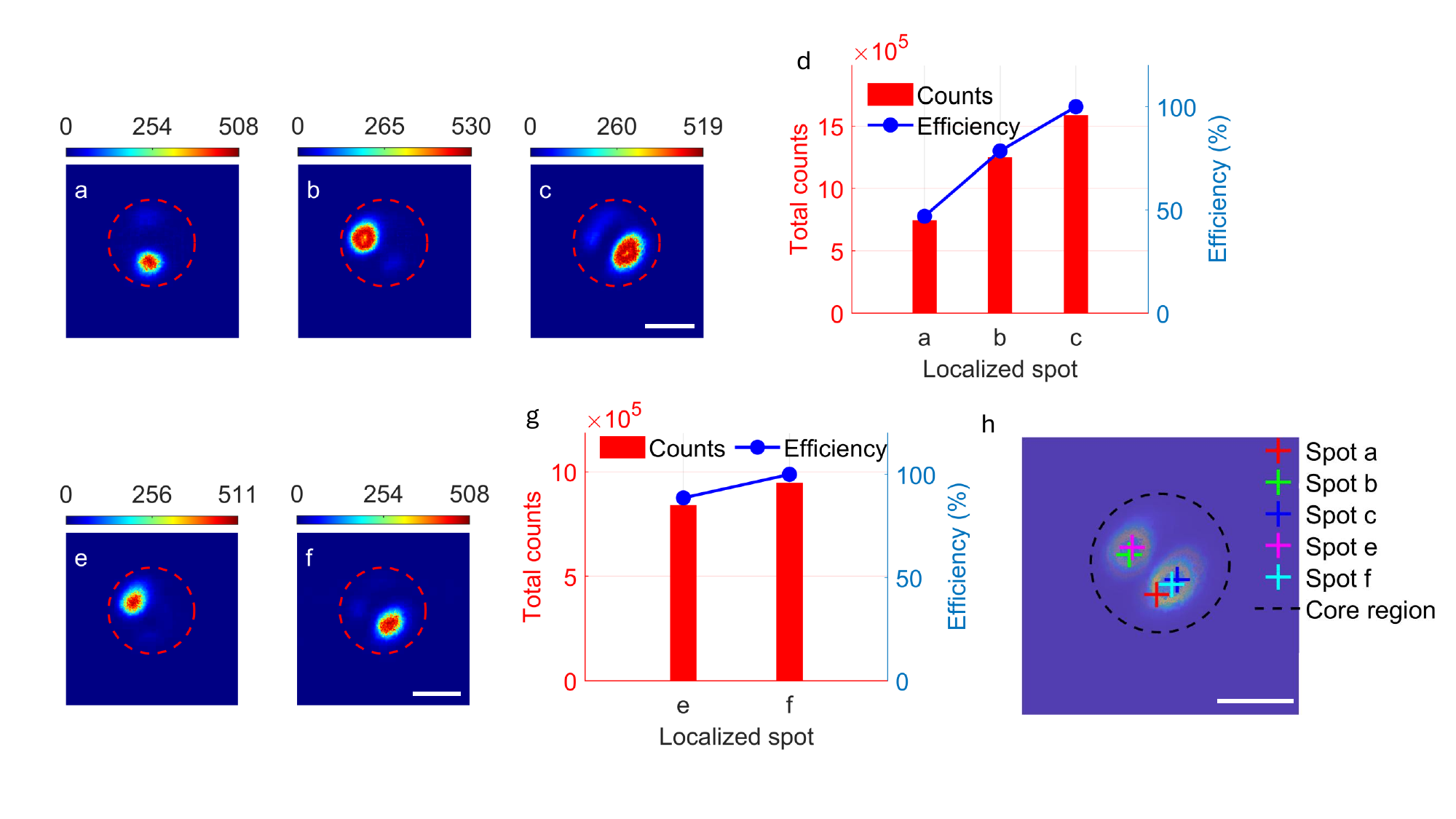}}
		\caption{Phase-locked mode localization under channel impairment. (a–c) Measured near-field intensity distributions at the MM facet of the three-mode PL with all three input ports active, showing three distinct Gaussian-like localized spots at different core regions. (d) Total detected photon counts and relative conversion efficiencies for the three localized modes, confirming efficient phase-locked recombination. (e–g) Corresponding measurements when one output port (PLO3) is deliberately blocked, simulating loss of an optical channel or actuator. Despite the reduced total counts ($\approx$40 \% of the three-port case), two bright, well-confined spots remain localized, demonstrating graceful degradation and stable mode formation. (h) Overlaid centroid positions of all five localized spots extracted from the three- and two-port configurations. The location corresponding to spot (a) is absent in the two-port case, confirming that channel loss limits the number of accessible positions but preserves localization stability and beam quality. Scale bars, 0.5 mm.}  
	\label{fig6}
	\end{figure*}
         
\noindent\textbf{Robust spot localization under channel impairment:} Robustness to channel impairment is critical for deployable MM systems, where connector drift, modal imbalance, or loss of a spatial channel can destabilize conventional adaptive-optics approaches. Our scheme exhibits graceful degradation under such impairments: disabling one lantern port reduces the number of accessible localized states, yet the remaining states retain their stability, localization quality, and high relative efficiency. To demonstrate this fault tolerance, we deliberately blocked one PL output and performed a full phase sweep. As presented in Figure~\ref{fig6}, even with a missing channel, the system reproducibly generated localized Gaussian-like spots at distinct positions on the MM facet, confirming that coherent recombination and spatial localization remain robust even when a degree of freedom is removed.

For this set of measurements, a different QWP angle was selected, yielding relative output intensities of 26\%, 40\%, and 34\% for PLO1, PLO2, and PLO3, respectively. When all ports were active, three well-confined spots appeared at distinct locations across the MM core, corresponding to spots a, b, and c, with conversion efficiencies of 47\%, 79\%, and 100\%, respectively. In other words, 74\% of the higher-order modal power from PLO2 and PLO3 was efficiently concentrated into a single localized spot—leaving negligible power outside that region—along with the remaining 26\% carried by the fundamental mode. Remarkably, when one output (PLO3) was blocked at the free-space delay line, two localized spots were still generated with the remaining $N-1$ ports, achieving efficiencies of 89\% and 100\% despite an overall count reduction of $\approx$40\% (matches well with the $\approx$1 dB insertion loss of PLO3). These results confirm that the loss of a single optical channel limits the number of accessible positions but does not compromise localization stability, efficiency, or polarization insensitivity.\\

\noindent\textbf{Phase-locked spot stability:} In fiber mode recombination systems, maintaining a stable output profile is often limited by environmental factors that can alter the relative modal phases and launch conditions. In our system, once the recombined spot was phase-locked, the output pattern remained highly stable over time without active feedback. To probe the resilience of the phase-locked output, we conducted two stability tests: a static measurement and a perturbation test in which the input fiber was gently agitated. Mechanical agitation induces uncontrolled higher-order mode coupling and random phase variations, serving as a stringent robustness benchmark for the recombination scheme. Under static conditions, we recorded a 44-s video at 25 fps (1104 frames), providing a high-resolution trace of any spatiotemporal drift in the localized output spot. The locked output maintained a consistent spatial distribution throughout the observation period, indicating that the established phase relationship between the modes is inherently robust under normal laboratory conditions.

\begin{figure*}[b]		
\centering{\includegraphics[width=\linewidth]{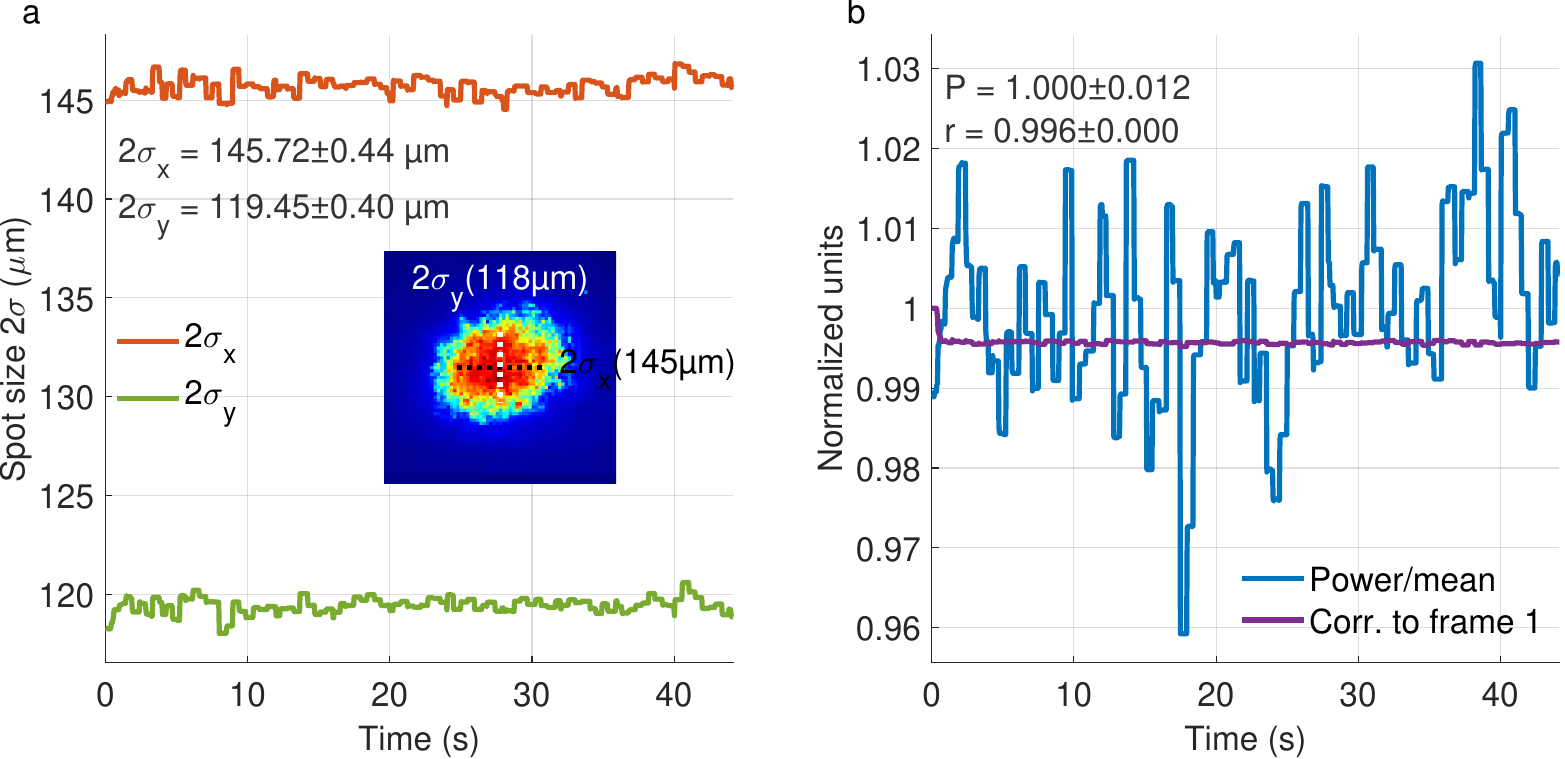}}
		\caption{Spatial and modal stability of the phase-locked spot. (a) Temporal evolution of the spot size during a 44-s acquisition. The inset shows frame 1 of the recorded video representing the generated spot, with the measured spot widths ($2\sigma_x = 145~\mu\mathrm{m},\; 2\sigma_y = 118~\mu\mathrm{m}$) indicated by dotted black and white lines along the respective axes. The mean spot widths and the standard deviation of the mean are denoted. (b) Temporal stability of the normalized power (blue) and frame-to-frame correlation relative to frame 1 (purple). The blue trace represents the total power in each frame, normalized to the mean power across the full acquisition. Power was obtained by summing the pixel intensities (photon counts) within the region of interest. Both the beam width and correlation remained highly stable throughout the measurement, as denoted in the respective plots.}  
	\label{fig7}
	\end{figure*}

As shown in Figure~\ref{fig7}, the phase-locked profile exhibited high stability in both spot size and power throughout the acquisition. The centroid, defined as the intensity-weighted mean position, was tracked with sub-pixel precision and converted to micrometers based on the $5.5~\mu\mathrm{m}$ pixel size. The beam size was quantified using the $1/e^2$ radii, $w_x=2\sigma_x$ and $w_y=2\sigma_y$, derived from the second intensity moments about the centroid. The total power was normalized to the mean across all frames, while the Pearson correlation coefficient was computed on mean-centered, variance-normalized intensity vectors—rendering it insensitive to global brightness fluctuations and primarily indicative of spatial profile stability.

 Across the acquisition, the radial RMS centroid drift is $0.553~\mu\mathrm{m}$, approximately $0.10$ pixels for a $5.5~\mu\mathrm{m}$ pitch, the spot widths are stable with $2\sigma_x = 145.718 \pm 0.444~\mu\mathrm{m}$ and $2\sigma_y = 119.449 \pm 0.396~\mu\mathrm{m}$, the normalized power is $1.000 \pm 0.012$, and the correlation remains $r = 0.996 \pm 0.000$, demonstrating a phase-locked interference state that is stable in position, size, and shape over tens of seconds. 

 The stability of the phase-locked profile was further assessed under changes in the input light phase and coupling conditions. In conventional fiber-based or free-space interferometric systems, small fluctuations in the input light phase or polarization can disrupt modal balance, causing beam wander and profile distortion. Such sensitivity arises because the input field directly defines the relative modal phases, making the output highly susceptible to environmental perturbations. In contrast, our PL architecture intrinsically suppresses input-induced phase distortions: the Faraday-reflected, phase-locked design ensures that the returning fields retrace identical optical paths, canceling polarization and phase fluctuations introduced at the input. To verify this robustness, the input light was continuously perturbed by bending and conforming the laser source fiber throughout the image acquisition, while the piezo-controlled phases were held fixed. A deliberately non-Gaussian mode was generated to enhance sensitivity to subtle shape variations under phase distortion.

 The video was recorded over a 181~s dataset, the generated phase-locked spot exhibited high stability with 
$\mathrm{RMS}_{\mathrm{drift}} = 3.14~\mu\mathrm{m}$, 
$2\sigma_x = 59.3 \pm 1.4~\mu\mathrm{m}$, 
$2\sigma_y = 80.6 \pm 0.6~\mu\mathrm{m}$, $p =1.000 \pm 0.015$, and 
$r = 0.990 \pm 0.006$. Together, these stability metrics show that, once phase-locked, the recombined spot remains effectively invariant over tens to hundreds of seconds: the centroid drifts by less than a pixel, the spot size remains constant, and the profile correlation stays near unity, even under deliberate input-phase disturbances. This highlights the exceptional robustness and phase-coherent character of our PL-based mode-recombination approach. \\

\noindent\textbf{Single-mode fiber coupling across output profiles:} Efficient coupling of structured beams into an SMF serves as a stringent benchmark of optical mode quality, as only the fundamental Gaussian mode of the fiber can be effectively excited. High SMF coupling efficiency not only reflects excellent spatial coherence and wavefront uniformity but is also crucial for applications requiring precise light delivery, interferometric stability, and low-loss integration into photonic systems. 

\begin{figure*}[t]	
\centering{\includegraphics[width=0.8\linewidth]{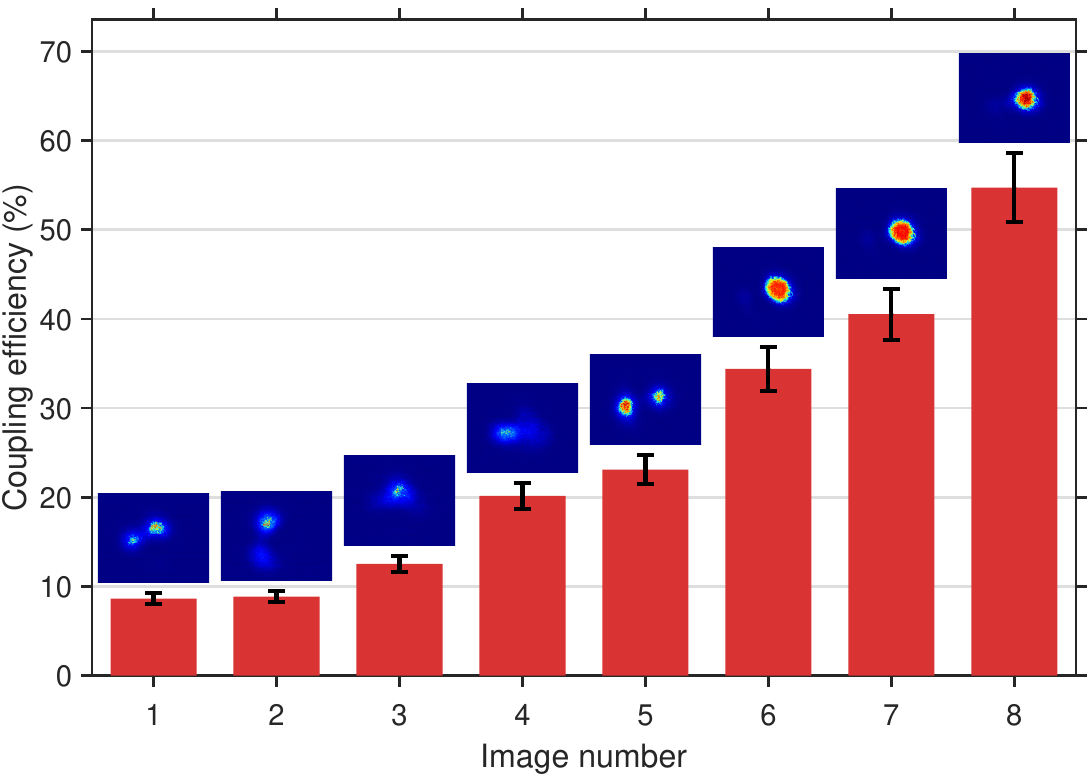}}
		\caption{SM coupling efficiency of the generated profiles with respective mode profile thumbnails. The highest efficiency was measured to be 55\%, corresponding to image 8.}
	\label{fig8}
	\end{figure*}
    
To quantify the modal purity and Gaussian similarity of the generated output profiles, we measured their coupling efficiency into a standard telecom single-mode fiber (SMF-28) (see Methods and Supplementary Note 4). Distinct spatial modes were first phase-locked by applying specific voltages to the PE stacks, ensuring reproducible output profiles. The reflected imaging arm was then routed through a coupling lens ($f=15~mm$) into the SMF, and the powers immediately before and after the fiber were recorded. The ratio of these values defines the SM coupling efficiency. The measured coupling efficiencies exhibited a strong dependence on beam quality: profiles with higher Gaussian fidelity yielded markedly higher SMF coupling. The best localized spot coupled 55\% of the incident power into the SMF with error bars corresponding to a measured uncertainty of $\pm$ 3.5\%, determined from repeated deliberate misalignment and realignment of the SMF coupling. The profile corresponds to the maximum efficiency shown in image 8, which also achieved a high Gaussian structural similarity index (Gaussian SSIM = 0.95). Figure~\ref{fig8} summarizes the trend, showing a bar plot of SMF coupling efficiency across all recorded profiles with overlaid representative images. The reported high SMF coupling values for the Gaussian-like profiles demonstrate that the generated modes not only overlap efficiently with the fiber’s fundamental profile but also meet the spatial-mode quality required for low-loss interfacing with downstream photonic and quantum-optical systems.

The prototype architecture shown in Fig. \ref{fig1} incorporates several auxiliary optical components that were included to provide maximum experimental flexibility and to validate the proof-of-concept for near-unity mode recombination. Because of this expanded component set, the end-to-end optical throughput of the current setup is measured to be –7.2 dB ($\approx$19 \% throughput). Importantly, this loss is dominated by testbed insertion losses—such as those from the PBS, fiber-pigtailed Faraday mirrors, and connector interfaces—rather than by the PL itself. A portion of the PL loss (3.4 dB bidirectional) also originates from coupling conditions at the MM input: if the launch NA or mode-field distribution does not match the acceptance profile of the tapered MM port, a fraction of the injected power is naturally filtered during the taper transition. These losses are implementation-specific and not intrinsic to the underlying recombination mechanism, and can be reduced through NA-matched launching or optimized beam sizing at the PL input.

A practical integrated design can use far fewer components and correspondingly much lower loss. For example, replacing the PBS with a slight-angle mirror configuration removes the beam splitter entirely from the optical path, and using compact free-space Faraday rotators instead of fiber-based reflectors reduces the return-path loss to below 0.5 dB. Together with improved NA-matched coupling to minimize the lantern insertion loss, these modifications reduce the overall system loss to below 2 dB while preserving the intrinsic near-unity recombination performance of the PL, and are readily scalable to low-loss, practical implementations.\\

 \section{Discussion}\label{sec3}

We demonstrate an all-fiber, polarization-independent scheme that enables real-time, reconfigurable control of the modal superposition at the MM input of a PL. Scanning the voltages applied to the integrated piezoelectric phase shifters allows access to a broad range of interference states without requiring external wavefront correction, transfer-matrix calibration, or polarization alignment. Importantly, the phase space contains discrete basins that correspond to stable, spatially localized field distributions, enabling controlled positioning of high-quality, Gaussian-like modes across the MM facet. This basin structure, together with the intrinsic reciprocity of the PL–FRM geometry, highlights the robustness and scalability of phase-only control for fiber-integrated beam shaping and signal processing.

By scanning only the $N-1$ relative phases in an $N$-mode lantern, the system accesses $N$ distinct phase basins, each generating a localized Gaussian-like spot at a different position on the MM facet, enabling deterministic and reconfigurable spatial routing. In contrast to adaptive-optics or SLM-based wavefront-shaping systems that rely on free-space propagation, polarization control, and continuous iterative optimization, the present PL–Faraday architecture achieves deterministic, phase-only multimode control in a fully guided-wave, polarization-reciprocal, and alignment-free platform. This reciprocal phase tuning directly enables stable deterministic modal interference for coherent beam combining, adaptive coupling, and reconfigurable beam delivery in integrated photonic systems.

The system maintains stable operation even under component loss, demonstrating inherent robustness. Localization by phase-locked mode recombination remains insensitive to input polarization and channel impairment due to the polarization-reciprocal return path and calibrated optical path lengths, requiring only relative phase locking of the active ports. This tolerance to optical and electronic faults—without polarization alignment or full actuator access—positions phase-locked MM recombination as a scalable, fault-tolerant platform for practical multimode photonics.

For deterministic spot localization, the three-port mode-selective PL provides an almost diagonal port–mode mapping (LP$_{01}$, LP$_{11a}$, LP$_{11b}$), allowing the $N-1$ controlled phases to translate directly into spatially localized spots with high extinction, near-unity efficiency, and reproducible behavior. This “calibration-light’’ operation enables a single coarse preset followed by a fine piezo scan to traverse multiple phase basins, while the fiber-integrated geometry and Faraday-mirror reciprocity ensure polarization-independent and environmentally stable operation. In contrast, conventional PLs implement general unitary port–mode maps that require full transfer-matrix calibration. The same phase-basin principle—accessing $N$ basins with $N-1$ phases—extends naturally to higher-mode devices, where additional mode-selective ports (e.g., 5–6 modes) broaden the set of accessible localized spots and enable richer spatial routing and beam forming (See Supplementary Note 7). Such higher-mode PLs can be readily driven by multi-channel FPGA-controlled piezoelectric phase shifters, offering a scalable path toward parallel MM control\cite{Lemons21}.

With conventional PLs containing multiple SM ports, our method remains applicable after a one-time transfer calibration. The complex transfer matrix $\boldsymbol{H}$ is measured from each SM port to the near-field output, and complex weights $\boldsymbol{a}$ maximizing overlap with a target Gaussian are computed. The corresponding phases are implemented via the piezo shifters (optionally with minor amplitude trimming through variable attenuators or split ratios), then stabilized using a slow dither-and-lock loop. Phase-only control already yields strong localization; modest amplitude balancing can further improve extinction. While transferable to generic PLs, mode-selective recombination offers a simpler, more reproducible, and higher-performance route for MM beam localization.

The measured stability further confirms the reliability of our system in maintaining phase-locked interference states. Under static conditions, the localized Gaussian-like mode exhibited sub-pixel drift and near-constant beam size and correlation over tens of seconds, while deliberate input-phase perturbations produced stable non-Gaussian modes without distortion or migration. These results demonstrate that the PL-based platform provides a robust, polarization-independent architecture for deterministic MM recombination with long-term spatial and modal stability. Such stability directly benefits coherence-critical applications where modal drift limits fidelity, including free-space quantum key distribution—where turbulence-induced phase distortion remains a major challenge \cite{Wang:20,articleaoQKD}—and astrophotonics, where deterministic modal shaping can enhance coupling efficiency and suppress modal noise in PL-fed spectrographs and astrophotometric instruments \cite{2012MNRAS,Lin:25}.

Beyond photonic communication and quantum domains, the demonstrated polarization-independent coherent recombination approach also holds promise for biomedical imaging and sensing. In optical coherence tomography (OCT)\cite{deSivry-Houle:21} and interferometric diagnostics\cite{HOSSEINZADEH2019111590,jbio.202100068}, multiple detection channels can be coherently combined through a PL to form an SM output, enhancing signal quality and SMF coupling efficiency while suppressing uncorrelated noise. The deterministic control and localization of Gaussian-like modes further enable compact, fiber-integrated beam-combining modules suitable for endoscopic or lab-on-fiber OCT systems. These attributes position the presented approach as a practical platform for noise-resilient, high-coherence biomedical interferometry and related diagnostic modalities.

\section*{Methods}\label{sec4}


\subsection*{Field synthesis at the MMF output}
The complex field at the MMF output is modeled as a coherent superposition of the fields launched by the $N$ PL arms\cite{Becerra-Deana:25,Kim_2024,Demur:23,s21196564,el:19970427}:

\begin{equation}
E(\mathbf{r}) \;=\; \sum_{m=1}^{N} c_m\,u_m(\mathbf{r})\,e^{i\phi_m},
\qquad \mathbf{r}=(x,y),
\label{eq:field-sum}
\end{equation}

Here \(u_m(\mathbf{r})\) denotes the normalized field distribution of port \(m\), and \(c_m\) includes the fixed amplitude and static phase offsets arising from coupling and propagation.

In practice, we realize separate single-spot states in distinct PE runs (scans) indexed by $k=1,2,3,\dots$ via a two-stage control:
\begin{equation}
\phi_m^{(k)} \;=\; \Phi_m^{(k)} \;+\; \delta\phi_m,
\label{eq:coarse-fine}
\end{equation}
where $\Phi_m^{(k)}$ is a coarse phase preset set by translating the delay line of arm $m$ for scan $k$, and $\delta\phi_m$ are fine PE phases swept during that scan. The detected intensity is $I^{(k)}(\mathbf{r})=\bigl|E^{(k)}(\mathbf{r})\bigr|^2$ with
$E^{(k)}(\mathbf{r})=\sum_m c_m u_m(\mathbf{r}) e^{i\phi_m^{(k)}}$.
Global phase is unobservable; the effective phase degrees of freedom are $N-1$ per scan. See supplementary Note 6 for the simulated results showing an excellent match to the experimentally observed interference patterns.

\subsection*{Localized spot formation as constructive interference}
Constructive interference of the guided modes leads to spatially localized Gaussian-like spots within the MM core, corresponding to experimentally observed bright regions that resemble SM outputs.
For scan $k$, let $\mathbf{r}_0^{(k)}$ denote the location of the desired Gaussian-like spot. A sufficient phase choice for constructive interference at $\mathbf{r}_0^{(k)}$ is
\begin{equation}
\phi_m^{(k)\star} \;=\; -\arg\!\bigl(c_m\,u_m(\mathbf{r}_0^{(k)})\bigr) \;+\; \phi_0^{(k)},
\label{eq:phase-align}
\end{equation}
for some offset $\phi_0^{(k)}\!\in\!\mathbb{R}$, which maximizes
\begin{equation}
E^{(k)}(\mathbf{r}_0^{(k)}) \;=\; \sum_{m=1}^{N} \bigl|c_m\,u_m(\mathbf{r}_0^{(k)})\bigr|.
\end{equation}
Operationally, the coarse preset $\{\Phi_m^{(k)}\}_m$ biases the interference landscape so that a subsequent fine PE sweep $\{\delta\phi_m\}_m$ converges to a single-spot solution near $\mathbf{r}_0^{(k)}$. Thus, repeating the experiment with different coarse presets $k=1,2,3$ yields three distinct single-spot realizations at different MMF-plane locations, one per scan. See Supplementary Note 6 for the simulated spot generation at $N$ basins for an ideal and perturbed three-mode PL, and Supplementary Note 7 for the corresponding six-mode commercial PL simulations demonstrating scalability to higher modal orders.

\paragraph{Phase–delay relation for coarse presets.}
The coarse phase preset is introduced through controlled path-length offsets that establish the initial interference condition. 
A translation $\Delta z_m^{(k)}$ in the free-space delay line or a corresponding fiber segment produces a phase shift given by
\begin{equation}
\Delta\phi_m^{(k)} \;=\; \frac{2\pi}{\lambda}\,\Delta\mathrm{OPL}_m^{(k)},
\qquad
\Delta\mathrm{OPL}_m^{(k)} \!=\!
\begin{cases}
2\,\Delta z_m^{(k)}, & \text{free-space round trip},\\[2pt]
2\,n_g\,\Delta L_m^{(k)}, & \text{fiber round trip},
\end{cases}
\label{eq:phase-delay}
\end{equation}
where $n_g$ is the group refractive index of the fiber. 
In practice, millimeter-scale translations are sufficient to establish the coarse phase preset $\Phi_m^{(k)}$, moving the system into a region of phase space where the subsequent PE-fine sweep $\{\delta\phi_m\}$ converges to a single Gaussian-like spot at the MM output.

\subsection*{Upper bounds on the number of independent localized peaks}
Let $\mathcal{U}=\operatorname{span}\{u_1,\dots,u_N\}$ denote the accessible modal subspace of the MM core. 
For a single-phase scan (fixed $k$) under phase-only control, the number $K$ of simultaneously realizable, comparable-brightness foci is practically bounded by
\begin{equation}
K \;\lesssim\; N-1 \;\le\; \operatorname{rank}(\mathcal{U}),
\end{equation}
reflecting the fact that only $N-1$ independent phase degrees of freedom govern constructive interference. 
Across multiple scans with distinct coarse presets $\{\Phi_m^{(k)}\}$, the system can sequentially access different single-spot solutions, one per scan. 
Accordingly, the three localized spots observed experimentally correspond to three independent phase basins explored across separate coarse presets.

\subsection*{Relation to Gaussian-mode overlap and SMF coupling}
Let $I(\mathbf{r})$ denote the background-corrected near-field intensity at the MMF output, where $\mathbf{r}=(x,y)$. 
To quantify the modal confinement and Gaussian similarity, $I(\mathbf{r})$ is compared to an ideal elliptical Gaussian distribution,
\begin{align}
u &= \cos\theta\,(x-x_0) + \sin\theta\,(y-y_0),\\
v &= -\sin\theta\,(x-x_0) + \cos\theta\,(y-y_0),\\
G(\mathbf{r};x_0,y_0,\theta,w_M,w_m,A) &= A\,\exp\!\left[-2\!\left(\frac{u^2}{w_M^2}+\frac{v^2}{w_m^2}\right)\right],
\end{align}
where $(x_0,y_0)$ denotes the spot center, $\theta$ is the major-axis orientation, and $w_M,w_m$ are the $1/e^2$ intensity radii (beam waists) along the principal axes; $A$ defines the peak amplitude. 
The orientation and beam widths are extracted from the principal-axis moments and $1/e^2$ intensity crossings, and $(x_0,y_0)$ is chosen to maximize the overlap metric defined below.

The complex-amplitude (mode) overlap between the measured and reference fields is given by
\begin{equation}
\gamma \;=\; 
\frac{\displaystyle \int_{\mathcal{C}} \sqrt{I(\mathbf{r})}\,\sqrt{G(\mathbf{r})}\,d^2\mathbf{r}}
     {\displaystyle \Bigl(\int_{\mathcal{C}} I(\mathbf{r})\,d^2\mathbf{r}\Bigr)^{1/2}
      \Bigl(\int_{\mathcal{C}} G(\mathbf{r})\,d^2\mathbf{r}\Bigr)^{1/2}},
\qquad
\eta \;=\; \gamma^2,
\label{eq:mode-overlap}
\end{equation}
where $\eta$ represents the normalized intensity (power) overlap, and the spot center $(x_0,y_0)$ is determined by
\begin{equation}
(x_0^\star,y_0^\star) \;=\; \arg\max_{x_0,y_0}\;\eta.
\end{equation}

In practice, the integration is confined to an adaptive, spot-centered elliptical window,
\begin{equation}
\mathcal{C} \;=\; \left\{(x,y): \left(\frac{u}{w_M}\right)^{\!2} + \left(\frac{v}{w_m}\right)^{\!2} \le 4 \right\},
\end{equation}
which focuses the metric on the localized bright region and suppresses background noise outside the spot. 
For weak aberrations and proper polarization alignment, $\eta$ provides a quantitative surrogate for the coupling efficiency into a matched SMF, since the coupled power into the SMF fundamental mode scales directly with $\eta$.

\section*{Additional Information}
See the Supplementary Material for more information.\\

\section*{Acknowledgments}

The authors thank Innovate UK (10002685); Royal Academy of Engineering RF\textbackslash201718\textbackslash1746); Engineering and Physical Sciences Research Council (EP/T001011/1); and Innovate-UK (10005967) for funding support. 

\section*{Declarations}
\subsection*{Conflict of Interest}
The authors declare no conflict of interest.

\section*{Author contribution}
HKC built and characterized the experimental setups, performed the measurements, performed theoretical modeling and simulations, designed the analysis program, analyzed the data, and consolidated the results. RD conceived the project and acquired funding. HKC wrote the manuscript and was reviewed by RD.



\clearpage
\appendix
\includepdf[pages=-]{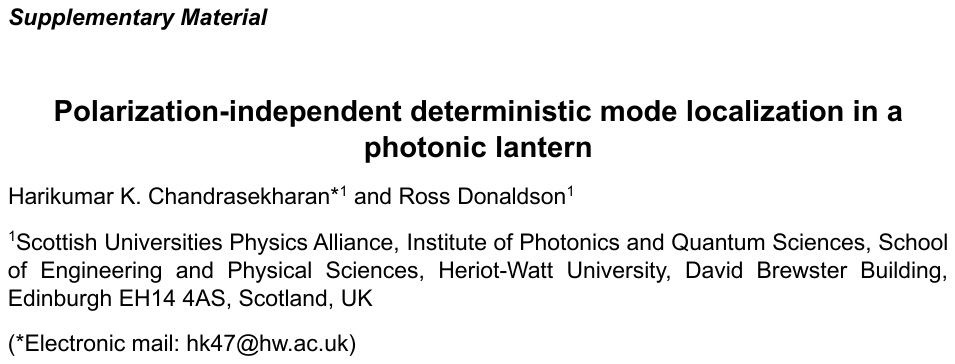}
\end{document}